\RequirePackage{fix-cm}
\documentclass[smallextended]{svjour3}       
\smartqed  
\usepackage{graphicx}

\usepackage{amssymb}
\usepackage{amsmath}
\begin{document}

\title{Coordinate effect: Vaidya solutions without integrating the field equations}



\author{E.G. Mychelkin \and M.A. Makukov} 


\institute{E.G. Mychelkin \at Fesenkov Astrophysical Institute\\
	\email{mychelkin@aphi.kz}           
           \and
           M.A. Makukov \at Fesenkov Astrophysical Institute\\
           \email{makukov@aphi.kz} 
}
\date{Received: date / Accepted: date}

\maketitle

\begin{abstract}
We extend Vaidya's algorithm for the description of a central mass losing or gaining energy due to electromagnetic-type radiation (`null dust') to the case of arbitrary radial corpuscular radiation. We also demonstrate the remarkable possibility of purely algebraic deduction of the Vaidya solution without integrating the field equations, and interpret this possibility as an artifact of curvature coordinates. Since Vaidya's approach by itself cannot lead to certain dependence of mass on spacetime coordinates, the search for a corresponding mass-function represents an independent issue. In this regard, as a perspective, we discuss an outlook on the problem of variable masses as a whole.

\end{abstract}

\section{Introduction}
\label{intro}
In general relativity (GR), the curvature coordinates play a special role.
Indeed, for example, in the case of spherical symmetry and given the interval
\begin{equation}
d{s}^2 = e^{\nu(r)} d{t}^2 - e^{\lambda(r)} d{r}^2 - r^2 d\Omega^2, 
\label{statcurvcoord}
\end{equation}
the angular part of the metric (the unit sphere) remains flat, greatly
simplifying the calculations. Moreover, in case of mutual reciprocity  of
the first two metric coefficients, $g_{00}g_{11}=-1$, as is the case for the
Schwarzschild solution,
\begin{equation}
d{s}^2 = \left(1-\frac{2m}{r}\right)d{t}^2 - \left(1-\frac{2m}{r}\right)^{-1} d{r}^2 - r^2 d\Omega^2,
\label{schwarz}
\end{equation}
the Einstein equations become linear \cite{Padmanabhan2010}.

As opposed to the Schwarzschild solution (\ref{schwarz}), in Vaidya's approach a mass, initially in vacuum,
begins to emit (or absorb) the radiation, which, in turn,
gravitates, i.e. is described by its energy-momentum tensor at the right side of
Einstein's equations. In a sense, we are dealing here with the extension of the
Schwarzschild problem to the case of a variable radiating mass, therefore, 
Vaidya's choice to search for the solution in curvature
coordinates is reasonable. A nontrivial moment in this technique is the
automatic inclusion of mass variability itself (the ``mass function'') into the
equations as a function of all 4-coordinates.

Here, we will show the remarkable possibility of a purely
algebraic derivation of the Vaidya metric solely on the basis of the Vaidya ansatz connecting the Schwarzschild seed solution with the general form of the solution in curvature coordinates.

Next, we demonstrate the universal nature of the Vaidya approach based on the
application of the curvature coordinates using the mechanism of mass loss due to
corpuscular radiation (e.g., stellar wind, massive neutrinos, etc.).

After discussing the peculiarities of using the curvature coordinates (primarily
as opposed to isotropic coordinates), we proceed to consider technical
as well as fundamental issues connected with the so-called first integral and
specific conditions for the mass-function necessary for
closing the system and compatibility check of each particular solution.

\section{Symmetries in the Vaidya problem}
The general relativistic problem of radiating mass was presented in the
pioneering works of Vaidya \cite{Vaidya1943,Vaidya1951,Vaidya1953}. The corresponding Einstein equations with radial emission (`null dust'),
\begin{equation}
	{G_{\mu}}^{\nu} \equiv {R_\mu}^\nu -\frac{1}{2}R \delta_\mu^\nu = \varkappa {T_{\mu}}^{\nu} = \varkappa \varepsilon k_\mu k^\nu, 
	\qquad 
	k_\alpha k^\alpha = 0, \quad k^2=k^3 = 0,
	\label{Ee}
\end{equation}
imply the following two types of symmetry conditions
\begin{equation}
	{T^{\mu}}_{\alpha} k^\alpha =0, \qquad \mu=0 \quad \Longrightarrow \quad {G_0}^0+{G_1}^0 e^{\frac{\nu-\lambda}{2}}=0,
	\label{e1}
\end{equation}
\begin{equation}
	{T_{\alpha}}^\alpha =0 \quad \Longrightarrow \quad {G_0}^0+{G_1}^1=0, 
	\label{e2}
\end{equation}
which are in non-static curvature coordinates
\begin{equation}
	d{s}^2 = e^{\nu(r, t)} d{t}^2 - e^{\lambda(r, t)} d{r}^2 - r^2 d\Omega^2 
	\label{curvcoord}
\end{equation}
take the form:
\begin{equation}
	e^{-\lambda}\left( \lambda^\prime - \frac{1}{r} \right) + \frac{1}{r} + \dot{\lambda} e^{-(\lambda+\nu)/2} = 0,
	\label{eo1}
\end{equation}
\begin{equation}
	e^{-\lambda}\left( \lambda^\prime - \nu^\prime - \frac{2}{r} \right) + \frac{2}{r} = 0,
	\label{eo2}
\end{equation}
where the prime denotes differentiation with respect to $ r $, and the dot with respect to $ t $. The third condition of symmetry, taken into account by Vaidya,
\begin{equation}
	{T_2}^2 = {T_3}^3 =0 \quad \Longrightarrow \quad {G_2}^2 = {G_3}^3 =0,
	\label{e3}
\end{equation}
or, explicitly,
\begin{equation}
	e^{-\lambda}\left( \frac{\lambda' - \nu'}{r} -\nu'' - \frac{{\nu'}^2}{2} + \frac{\lambda'\nu'}{2}   \right) 
	+ e^{-\nu} \left( \ddot{\lambda} + \frac{\dot{\lambda}^2}{2} - \frac{\dot{\lambda}\dot{\nu}}{2}  \right) = 0,
	\label{eo3}
\end{equation}
is in fact not independent. Indeed, given the zero trace of the energy-momentum
tensor, we have from \eqref{e3}:
\begin{equation}
	{R_2}^2 = {R_3}^3 = 0,
	\label{e3a}
\end{equation}
so that (\ref{eo3}) reduces to (\ref{e3a}), which, in its turn, appears to be equivalent (up to a constant factor) to the equation (\ref{eo2}).

The key point is the main Vaidya's ansatz according to which, for
reasons of correct asymptotics at infinity, it is assumed that the $ g_{11}
$-component of the metric (\ref{curvcoord}) is represented in the
Schwarzschild form, but with the replacement of the constant mass by a certain
mass-function $ m (r, t) $ \cite{Vaidya1951}:
\begin{equation}
-g_{11}=e^{\lambda(r,t)}=\left( 1-\frac{2m(r,t)}{r} \right)^{-1}\equiv D^{-1}.
\label{ansatz}
\end{equation}
Then the problem reduces to determining the component of the metric $ g_ {00} $, which, as it turned out, in the curvature coordinates can be found, up to an unknown mass-function $ m (r, t) $, algebraically from the condition (\ref{e1}).

\section{Vaidya's solution as an algebraic consequence of the main ansatz}
Indeed, the condition (\ref{e1}) in the form (\ref{eo1}) is an algebraic relation that can be easily solved with respect to $ g_ {00} $:
\begin{equation}
g_{00} = e^{\nu(r,t)} = \frac{\dot{\lambda}^2 e^{\lambda}r^2}{\left( e^{\lambda} + \lambda^\prime r - 1\right)^2 }.
\label{enu}
\end{equation}
Substituting the ansatz (\ref{ansatz}) instead of $e^{\lambda}$, we finally get:
\begin{equation}
e^\nu = \frac{\dot{m}^2}{{m^\prime}^2}\left( 1 - \frac{2m(r,t)}{r}\right)^{-1}= \frac{\dot{m}^2}{{m^\prime}^2}D^{-1}.
\label{enu2}
\end{equation}
Alternatively, following \cite{Vaidya1951}, we substitute (\ref{ansatz}) directly into (\ref{eo1}), which gives:
\begin{equation}
{m^\prime} e^{\nu/2} =- \dot{m} e^{\lambda/2}.
\label{funcrel}
\end{equation}
This relation is equivalent to (\ref{enu2}). In both cases, we get the Vaidya solution:
\begin{equation}                                                        
d{s}^2 = \frac{\dot{m}^2}{{m^\prime}^2} D^{-1}d{t}^2 - D^{-1} d{r}^2 - r^2 d\Omega^2.
\label{vaidyametric}
\end{equation}
So, due to radiation symmetries expressed in curvature coordinates as (\ref{eo1}, \ref{eo2}, \ref{eo3}), two unknown
functions $ \lambda (r, t) $ and $ \nu (r, t) $ are reduced with the described
method to a single one, $ m = m (r, t) $, without necessity of integration of
the field equations (for the traditional approach see, for example,
\cite{Mychelkin1990a}) and without using the explicit form of the energy-momentum tensor on the right side. The energy density being necessary for self-consistency of the task is established after substituting the metric (\ref{vaidyametric}) in the field equations, which gives
\begin{equation}
	\varepsilon = \frac{m'}{4 \pi r^2},
\label{enden}
\end{equation}
or, in integral representation,
\begin{equation}
	m(r,t) \equiv \int_{0}^{r}4 \pi \varepsilon(t,r) r^2 dr.
	\label{int}
\end{equation}
The latter expression was used by Raychaudhuri \cite{Raychaudhuri1953} as the initial
ansatz for the subsequent alternative deduction of the Vaidya metric
(\ref{vaidyametric}). Importantly, from \eqref{int} one may see that the radiated
mass-energy is represented by a function of both time and radial coordinate simultaneously.

\section{Corpuscular radial emission}
\label{sec:nonnull}
Let us consider a direct extension of the Vaidya algorithm, which does not
require the integration of the Einstein equations as well, to the case of arbitrary
radial corpuscular radiation (including bradyon and tachyon variants); in part,
we follow here the unpublished preprint \cite{Mychelkin1990a}, while correcting its
inaccuracies and also modifying the interpretation of the results. Then, Einstein's equations (\ref{Ee}) can be generalized as follows:
\begin{equation}
	{G_{\mu}}^{\nu}\equiv {R_\mu}^\nu -\frac{1}{2}R \delta_\mu^\nu = \varkappa {T_{\mu}}^{\nu} = \varkappa \varepsilon k_\mu k^\nu, 
	\label{eenn}
\end{equation}
\begin{equation}
	k_\nu k^\nu = 
	\begin{cases}
		\,\,\,\, 1 \\
		\,\,\,\, 0 \\
		-1
	\end{cases}   , \quad  k^2=k^3 = 0,
	\label{EeG}
\end{equation}
which includes all situations: $ k_ \nu k ^ \nu = 0 $ (null-dust), $ k_
\nu k ^ \nu = 1 $ (bradyons) and $ k_ \nu k ^ \nu = -1 $ (tachyons). Each of
these conditions implies its own separate parameterization. In general, we adopt
\begin{equation}
	k^\mu = \alpha u^\mu + \beta n^\mu,
\end{equation}
$$
u_\nu u^\nu = 1, \quad n_\nu n^\nu = -1, \quad u_\nu n^\nu = 0,
$$
i.e.
\begin{equation}
	k_\nu k^\nu =  \alpha^2 - \beta^2.
\end{equation}
Thus, the ordinary case (null-dust) is parameterized as $ \alpha = \beta $, with equal contributions from time- and space-like parts.

First, we study the time-like version $ k_ \nu k ^ \nu = 1 $ (``bradyon dust
emission'') and accept the following gauge conditions for
the radial vector $ k ^ \mu $ of the collinear initial particle velocity $ 0 \le
v <1 $ (in units of the speed of light):
\begin{equation}
	k^\mu = \alpha u^\mu + \beta n^\mu= \frac{u^\mu}{\sqrt{1-v^2}} +
	\frac{v n^\mu}{\sqrt{1-v^2}}\,,
	\label{k}
\end{equation}
with basis unit vectors
\begin{equation}
	u^\mu = {\delta_0^\mu}/{\sqrt{g_{00}}}, \quad n^\mu = {\delta_1^\mu}/{\sqrt{-g_{11}}},
	\label{un} 
\end{equation}
for which
\begin{equation}
	k^0 = \frac{\alpha}{\sqrt{g_{00}}}, \, \, k^1 = \frac{\beta}{\sqrt{-g_{11}}}, \, \, k_0 = \alpha\sqrt{g_{00}}, \, \, k_1 = -\beta\sqrt{-g_{11}}.
	\label{kk}
\end{equation}
Then, the previous symmetry conditions (\ref{e1}) and (\ref{e2}) turn into
\begin{equation}
	{T_{\alpha}}^{\mu} k^\alpha = \varepsilon k^\mu , \quad \mu=0 \quad \Longrightarrow \quad {G_0}^0 k^0+{G_1}^0 k^1 = \varkappa\varepsilon k^0 
	\label{ee1G}
\end{equation}
and
\begin{equation}
	{T_{\alpha}}^\alpha = \varepsilon \quad \Longrightarrow \quad {G_0}^0+{G_1}^1=\varkappa\varepsilon \, ,
	\label{ee2G}
\end{equation}
and the third condition (\ref{e3}) becomes:
\begin{equation}
	{T_2}^2 = {T_3}^3 =0 \quad \Longrightarrow \quad {G_2}^2 = {G_3}^3 =0 \quad \Longrightarrow \quad {R_2}^2 = {R_3}^3 = -\varkappa\varepsilon/2 \, .
	\label{ee3G}
\end{equation}
Now, unlike the null-dust case, from the equations (\ref{ee1G})-(\ref{ee3G})
one needs to exclude the intrinsic energy density $ \varepsilon $ using Einstein's equations, for example:
\begin{equation}
	{G_1}^0=\varkappa\varepsilon k_1 k^0 \qquad \Longrightarrow \qquad \varkappa\varepsilon k^0 = \frac{{G_1}^0}{k_1},
	\label{ee10}
\end{equation}
from which it also follows that
\begin{equation}   
\varepsilon=\frac{2}{\varkappa}\frac{m'}{\alpha^2 r^2} .
\label{epsil}
\end{equation} 
Then, substituting (\ref{ee10}) into (\ref{ee1G}) taking into account (\ref{kk}) and (\ref{curvcoord}) we finally get in curvature coordinates:
\begin{equation}
	{G_0}^0 - \frac{k_0}{k_1}{G_1}^0 = {G_0}^0 + \frac{\alpha\sqrt{g_{00}}}{\beta\sqrt{-g_{11}}}{G_1}^0 = {G_0}^0 + \frac{\alpha}{\beta}e^{\frac{\nu-\lambda}{2}}{G_1}^0 = 0,
	\label{ee1GG}
\end{equation}
where
$$ {G_0}^0=\frac{e^{-\lambda}}{r^2}\left(  r\lambda' + e^\lambda -1  \right), \qquad {G_1}^0=\frac{e^{-\nu}\dot{\lambda}}{r}. $$
Applying the Vaidya ansatz (\ref{ansatz}) to (\ref{ee1GG}) after transformations, we find an analogue of the relation (\ref{funcrel}):
\begin{equation}
	\beta m' e^{\nu/2} = -\alpha \dot{m} e^{\lambda/2},
	\label{eqG}
\end{equation}
or, in the final form,
\begin{equation}
	g_{00} = e^\nu = \frac{\alpha^2}{\beta^2}\frac{\dot{m}^2}{{m'}^2} \left( 1-\frac{2m}{r}  \right)^{-1} = \frac{\alpha^2}{\beta^2}\frac{\dot{m}^2}{{m'}^2} D^{-1}=\frac{1}{v^2}\frac{\dot{m}^2}{{m'}^2} D^{-1},
	\label{eqG00}
\end{equation}
where again $ m $ stands for $ m (r, t) $.

Note that for $ \beta / \alpha = v \rightarrow 0 $ from (\ref{eqG}) it follows $
\dot{m} \rightarrow 0 $ (as it should be from physical considerations), so $v
\rightarrow 0 $ in the expression (\ref{eqG00}) does not necessarily imply divergence.

Thus, the Vaidya metric for generalized (corpuscular) radial emission,
\begin{equation}                                                        
	d{s}^2 = \frac{\alpha^2}{\beta^2} \frac{\dot{m}^2}{{m^\prime}^2} D^{-1}d{t}^2 - D^{-1} d{r}^2 - r^2 d\Omega^2
	\label{vaidyametricG},
\end{equation}
is obtained again without integrating the field equations, due to the same radial symmetries in curvature coordinates.
The considered non-null emission could also be represented as a source for the
Vaidya-type metrics with non-zero radial pressure \cite{Culetu2016}.

As for the space-like sector (tachyon
dust emission, compare with \cite{Foster1972}) defined by the requirement $ k_ \nu k ^ \nu = -1 $, analytically
it differs from the one considered above only in the initial parameterization in
the gauge condition of the type (\ref{k}):
\begin{equation}
	k^\mu = \alpha u^\mu + \beta n^\mu= \frac{u^\mu}{\sqrt{v^2-1}} + \frac{v n^\mu}{\sqrt{v^2-1}}, 
	\label{kt}
\end{equation}
where $ 1 <v <\infty $ in this gauge. Such tachyon emission might be of
interest, as modern experiment is not capable of excluding the
possibility of the existence of background neutrinos in tachyonic state (see,
e.g., \cite{Caban2006,Ehrlich2015,Mychelkin2015,Makukov2016}).

\section{Cosmological term in curvature and isotropic coordinates}
\label{sec:Lam}
The described Vaidya algorithm can be extended to the case of the simplest
cosmological background \cite{Mallett1985} if we include the regular $ \Lambda $-term into the ansatz
(\ref{ansatz}), similar to the case of the Kottler
(Schwarzschild--de Sitter) metric
\begin{equation}
-g_{11}=e^{\lambda(r,t)}=\left( 1-\frac{2m(r,t)}{r} - \frac{\Lambda r^2}{3} \right)^{-1}= \tilde{D}^{-1}.
\label{ansatzdes}
\end{equation}
After substituting into the extended condition of the type (\ref{e1}),
\begin{equation}
\varkappa{T_{\alpha}}^{\mu} l^\alpha =\Lambda \delta_0^\mu \quad \Longrightarrow \quad {G_0}^0+{G_1}^0 e^{\frac{\nu-\lambda}{2}} - \Lambda=0,
\label{ee1}
\end{equation}
algebraic transformations lead to Vaidya-type metric in the curvature
coordinates embedded into the de Sitter background, similar to (\ref{enu}),
(\ref{enu2}) and (\ref{vaidyametric}) with the only replacement of $ D $ by $ \tilde{D} $:
\begin{equation}                                                        
d{s}^2 = \frac{\dot{m}^2}{{m^\prime}^2}\tilde{D}^{-1}d{t}^2 - \tilde{D}^{-1} d{r}^2 - r^2 d\Omega^2,
\label{vaidyades}
\end{equation}
and again without integration of the Einstein equations (in curvature
coordinates only).

To see how much the situation changes by considering isotropic coordinates
instead,
\begin{equation}
	d{s}^2 = e^{\nu(R, t)} d{t}^2 - e^{\mu(R, t)} \left(   d{R}^2 + R^2
	d\Omega^2 \right) ,
	\label{isotropcoord}
\end{equation}
we first consider the case of constant mass in the
definition of $\tilde{D}$ (\ref{ansatzdes}), i.e. the standard Kottler
metric in curvature coordinates (for which, as in the Schwarzschild case,
$g_{00}g_{11}=-1$). Even in this simplest case deducing the transformation from
curvature to isotropic coordinates is nontrivial, and was first obtained
in \cite{Kozhanov1986b}. Since this result is interesting in
itself and is widely unknown, we briefly outline it here. Using the standard
Tolman algorithm \cite{Tolman1949}, we get from the Kottler metric:
\begin{equation}
\int\frac{dR}{R}=\int\frac{dr}{r} e^{{\lambda(r)}/{2}}=\int\frac{dr}{\sqrt{(-{\Lambda}/{3})r^4 + r^2-2mr}} .
\label{elliptic}
\end{equation}
Via the substitution found by the polynomial method from 
\cite{Zhuravsky1941a}
\begin{equation}
r=\frac{6m}{1-12y}=-\frac{m}{2 \left( y-\frac{1}{12} \right)},
\label{D3}
\end{equation}
the elliptic integral \eqref{elliptic} is reduced to the Weierstrass canonical form
\begin{equation}
\int_{x_0}^{x} dx = \int_{\ln R_0}^{\ln R} d\ln R = \int_{y_0}^{y}\frac{dy}{\sqrt{4y^3-g_2 y-g_3}},
\end{equation}
whose inversion is the elliptic Weierstrass function with invariants $g_2$ and
$g_3$,
\begin{equation}
y=\wp(x,g_2,g_3), \quad x = \ln{\frac{R}{R_0}}, \quad g_2=\frac{1}{12}, \quad g_3=\frac{\Lambda m^2}{12} - \frac{1}{6^3}.
\label{D5}
\end{equation}
Therefore, according to (\ref{D3}) and (\ref{D5}), the transformation
from curvature to isotropic coordinates is:
\begin{equation}
r(R) = -\frac{m}{2(\wp-1/12)}.
\end{equation}
Hence, the interval for the Kottler metric in isotropic coordinates, after
some algebra \cite{Kozhanov1986b}, assumes the final form:
\begin{equation}
ds^2 = \frac{\wp'^2}{(\wp - 1/12)^2}dt^2-\frac{m^2}{4R^2(\wp - 1/12)^2}\left(dR^2 + R^2 d\Omega^2  \right).
\label{kottlerisotrop}
\end{equation}
Now, $\Lambda$-term enters the functions $\wp$ and
$\wp'=d\wp(x,g_2,g_3)/dx$ through
the parameter $g_3$, in accord with \eqref{D5}.

However, moving in (\ref{elliptic}) to general Vaidya case, when
$\lambda(r)\rightarrow \lambda(r,t)$, we get, instead of the constant $m$, an
unknown mass function $m(r,t)$. In such case, integration and, accordingly,
conversion to isotropic coordinates becomes impossible.

\section{Relation of Vaidya algorithm to isotropic and null coordinates}
\label{sec:AppendixC}
Although there is no explicit transformation of the Vaidya metric with an
unknown mass-function to other frames, we can still try to apply the Vaidya
algorithm directly in isotropic coordinates
\begin{equation}
d{s}^2 = e^{\nu(R,t)} d{t}^2 - e^{\mu(R,t)} \left( d{R}^2 + R^2 d\Omega^2 \right).
\end{equation}
In such a case, since the null-dust symmetries (\ref{e1}), (\ref{e2}) and (\ref{e3}) are preserved, the field
equations corresponding to (\ref{eo1}), (\ref{eo2}) and (\ref{eo3}) will be as follows:  
\begin{equation}
\frac{e^{\nu/2}}{R} \left[ \mu' \left(R\mu' + 8 \right) +4 R \mu'' \right] +2 e^{\nu-\mu/2}\left( \dot{\mu}\nu' -2 \dot{\mu}' \right) - 3 \dot{\mu}^2 e^{\mu-{\nu}/{2}} = 0,
\label{2b}
\end{equation}
\begin{equation}
\frac{e^{-\mu}}{R} \left[ \mu' \left( R \mu' + R \nu' + 6 \right) + 2R\mu'' + 2\nu' \right] + e^{-\nu} \left[ \dot{\mu} \left( \dot{\nu}-3\dot{\mu} \right) -2 \ddot{\mu} \right]  = 0,
\end{equation}
and
\begin{equation}
2\left(\mu'+\nu'\right) + R \left[ \nu'^2 + 2\left( \mu'' + \nu'' \right)  + e^{\mu-\nu} \left( 2\dot{\mu}\dot{\nu} - 3\dot{\mu}^2 - 4\ddot{\mu} \right)  \right] = 0,
\label{4b}
\end{equation}
where the derivatives with respect to $R$ are now denoted as primes. Further, following
Vaidya's algorithm, we again consider the Schwarzschild metric as an initial reference, but
in isotropic, rather than curvature, coordinates,
\begin{equation}
ds^2=\left(  \frac{1-m/2R}{1+m/2R} \right)^2dt^2 - \left( 1+\frac{m}{2R}  \right)^4 (dR^2 +R^2 d\Omega^2).
\end{equation}
Based on this, we adopt the `isotropic" ansatz as direct analog of (\ref{ansatz}): 
\begin{equation}
e^{\mu(R,t)} = \left(  1+\frac{m(R,t)}{2R}  \right)^4 \equiv B^4(R,t),
\label{grrisotr}
\end{equation}
where the constant mass is replaced by the mass-function $m=m(R, t)$. 

However, unlike in curvature coordinates, the remaining component of the desired isotropic metric $g_{00} (R,t)$ does not follow algebraically from the corresponding equation (\ref{2b}). Instead, substituting (\ref{grrisotr}) in (\ref{2b}), we arrive at the following differential equation with respect to $g_{00} = e^\nu$:
\begin{equation*}
y'+\frac{3y}{R B} - \frac{9\dot{m}}{2 R}B^5 = 0,
\end{equation*}
where $y=e^{3\nu/2}$.
The solution of this equation (and of the whole system (\ref{2b})-(\ref{4b}))
cannot be obtained if the explicit dependence $m=m (R,t)$ is unknown. Thus, the
Vaidya algorithm does not work in isotropic coordinates, and the transformation
from curvature to isotropic coordinates does not exist either.

By direct analogy with the Schwarzschild case, there is another commonly used representation of the Vaidya metric in the
Eddington-Finkelstein null coordinates \cite{Vaidya1953}, namely:
\begin{equation}
ds^2=\left( 1-\frac{2 m(u)}{r} \right)du^2 \pm 2du dr - r^2 d\Omega^2.
\label{EF}
\end{equation}
Expression (\ref{EF}) differs from its Schwarzschild counterpart only by the replacement of 
$m=\text{const}$ with the variable mass function $m(u)$. But now the transfer
from curvature to null  coordinates deserves a deeper consideration. Traditionally, the metric (\ref{EF}) can be
obtained from the original Vaidya metric (\ref{vaidyametric}) using the differential transformation 
\begin{equation}
dt=du\pm \frac{dr}{1-\frac{2 m}{r}}=du \pm dr^*,
\label{EFtransform}
\end{equation}
where $r^*$ corresponds to the so-called ingoing ($+$) or outgoing ($ - $)
tortoise coordinates. For the Schwarzschild problem in standard curvature
coordinates this transformation represents the usual way to go to the metric
(\ref{EF}) with $m=\text{const}$. Before applying this transformation to the variable
$m \ne \text{const}$, one must first represent the factor $\dot{m}^2/m'^2$ in (\ref{vaidyametric}) as
\begin{equation}
\frac{\dot{m}^2}{m'^2}=\dot{r}^2 = \left( 1-\frac{2m(r,t)}{r} \right)^2,
\end{equation}
which corresponds to fixing the value of (\ref{EFtransform}) on the lines of
constant $u$. In this particular case, (\ref{vaidyametric}) takes the ``desired form'',
\begin{equation}
d{s}^2 = \left(1-\frac{2m(r,t)}{r}\right)d{t}^2 - \left(1-\frac{2m(r,t)}{r}\right)^{-1} d{r}^2 - r^2 d\Omega^2,
\label{desired}
\end{equation}
and then the metric (\ref{EF}) is obtained formally using the traditional algorithm
(\ref{EFtransform}). However, this transition to $m(u)$ is dubious, since in case of
an unspecified mass function $m=m(r,t)$, the expression (\ref{EFtransform})
cannot be integrated. In other words, the final representation of the trailing (leading) coordinate $u = t \pm r^*$ is defined only up to the general quadrature
$$r^* = \int \frac{dr}{1-\frac{2m}{r}}\,,$$
which, in this case, cannot be explicitly integrated (see also in
\cite{Culetu2016}).
Thus, the Vaidya metric in null coordinates (\ref{EF}) appears, in a sense, symbolic rather than operational. 

Summing up the last two sections, one might say that when searching for explicit  
transformations of Vaidya's solution from the curvature coordinates, one
always has to know the explicit form of the mass function.

\section{First integral and mass-function}
The Einstein equations in the Vaidya metric can only be integrated up to the
so-called ``first integral''. This brings us to the discussion of the remaining
equation (\ref{eo2}). Substituting the metric components (\ref{ansatz}) and
(\ref{enu2}) into (\ref{eo2}), we arrive at an important identity:
\begin{equation}
\left(\frac{m''}{m'}-\frac{\dot{m}'}{\dot{m}}\right)D + D'+\frac{2m'}{r}=
\left(\frac{m''}{m'}-\frac{\dot{m}'}{\dot{m}}\right)D +\frac{2m}{r^2}=0.
\label{u01}
\end{equation}
From here, in particular, we can get the known first integral \cite{Vaidya1951}: 
\begin{equation}
f(m) = m^\prime \left( 1 - 2m/r \right).
\label{firstint}
\end{equation}
Substitution of (\ref{firstint}) into (\ref{vaidyametric}) leads to another commonly used representation of the Vaidya metric: 
\begin{equation}
d{s}^2 = \frac{\dot{m}^2}{f^2(m)} \left(1-\frac{2m}{r}\right)d{t}^2 - \left(1-\frac{2m}{r}\right)^{-1} d{r}^2 - r^2 d\Omega^2,
\label{vaidyametric2}
\end{equation}
where for brevity we write $m$ instead of $m(r,t)$.

Vaidya interpreted (\ref{firstint}) as a differential equation that should be
resolved with respect to $m$ \cite{Vaidya1951}. Formally, the formula
(\ref{firstint}) is a relation between two unknown functions $m(r,t)$ and $f(m)$
to be used for the definition of $m$. 

From the standpoint of searching for possible restrictions on the form of $f(m)$,
on the contrary, one can substitute a general metric (\ref{vaidyametric2})
directly into the Einstein equations (\ref{Ee}). Then, in particular, using the
$22$- and $01$-components, we get  
\begin{equation}
\left( \frac{\dot{m}^\prime}{\dot{m}} -\frac{f^\prime}{f} \right)D=\frac{2m^\prime}{r}, \quad D=1-\frac{2m(r,t)}{r}\,.
\label{A1}
\end{equation}
Further, given that $f=f(m)$ \cite{Vaidya1951}, we can write $\dot{f}(m)=f_ {,
m}\dot{m}$ and, similarly, $f^\prime(m)=f_ {, m}m^\prime$ (thus  
$
\frac{f^\prime}{f}=\frac{\dot{f}}{f}\frac{m^\prime}{\dot{m}}
\label{A2}
$)
and multiplying now (\ref{A1}) by $\dot{m}/m^\prime$, one gets
\begin{equation}
\frac{\dot{f}}{f}-\frac{\dot{m}^\prime}{m^\prime}+\frac{2\dot{m}}{rD}=0.
\label{A3}
\end{equation}
But $2\dot{m}/r=-\dot{D}$, so by integrating (\ref{A3}), we get instead of
(\ref{firstint}) some more general valid form of the first integral, where the role of the integration constant $C$ is played, generally speaking, by an arbitrary function of the radial coordinate $r$:
\begin{equation}
\left( \ln \frac{f}{m^\prime D}  \right)^{\boldsymbol{\cdot}}=0 \quad \Longrightarrow \quad f=\frac{m^\prime D}{C(r)}.
\label{ffull}
\end{equation}
In \cite{Gupta1988}, an expression similar to (\ref{ffull}) is obtained, and
arguments are made in favor of choosing $C(r)=1$, which, as expected, coincides
with the original functional (\ref{firstint}), and, in turn, actually turns
the Vaidya metric (\ref{vaidyametric2}) back to its original form (\ref{vaidyametric}) with the only unknown mass function $m=m(r,t)$.

\section{Discussion}
We have seen that for the presentation of Vaidya's solution in closed form in
curvature or null coordinates it is
necessary to know the explicit dependence of the mass-function on spacetime
coordinates, and this is a separate issue.
There were attempts to bring the metric in the form (\ref{vaidyametric2}) 
asymptotically to the standard Schwarzschild form (\ref{desired}) by
imposing a physically acceptable constraint of the type $f(m) = \pm \dot{m}$
\cite{Mallett1985,Vaidya1951,Gupta1988}. However, direct
substitution of the metric (\ref{desired}) into the field equations shows that although the condition (\ref{eo2}) is satisfied, the other equation (\ref{eo1}) is incompatible with the requirement $f(m) = \pm \dot{m} \ne 0$. 

In fact, this could be expected, because it follows from (\ref{funcrel}) and
(\ref{vaidyametric}) that the mass function $m(r, t)$ cannot be a function of
only $r$ or only $t$. In particular, this means that the known phenomenological
Eddington-Jeans power-law, $\dot{m}=-km^n$, frequently used to describe stellar
evolution, cannot be incorporated into the Vaidya approach.

We emphasize that, in any case, the freedom to choose the mass function $m(r,t)$
is limited by the functional relation (\ref{u01}), or its consequence
(\ref{A1}), which, however, is difficult to satisfy in practice. For
example, from (\ref{u01}) we can again obtain that the condition $m=m(t)$ is not compatible with the Vaidya metric, and the same is true for the simplest dependencies of both multiplicative $m(r,t)= f(r)\phi(t)$ and additive $m(r, t)=\alpha f(r) + \beta \phi(t)$ types.

In general, the specific nature of solutions involving the use of curvature
coordinates in GR is not an exception. Thus, the Fisher solution
\cite{Fisher1948} of the Einstein equations with minimal scalar field in curvature coordinates is
also found only up to a non-trivial (not analytically solvable) identity,
although in isotropic and other coordinates this problem is solved in exact form
\cite{Janis1968,Xanthopoulos1989} (for more details on this point, see
\cite{Makukov2020}). Besides, it is worth to note that the extension of the Kottler metric onto variable cosmological term is uniquely
related to curvature coordinates as well \cite{Dymnikova}. The curvature and
related null coordinates also play crucial role in deducing the Kerr rotational
solution \cite{Kerr1963,Newman1965} and in performing the Choptuik dynamical chaos
algorithm for collapsing scalar field \cite{Choptuik1993}.

There exist also more mathematical studies (e.g., \cite{Shaikh2019})
on geometric symmetries generated by the structure of the Riemann-Christoffel and Weyl conformal curvature tensors, without examining the mass-function properties\footnote{Acting in  this way, the authors in \cite{Shaikh2019} analyze the
pure radiation case in Vaidya's (null) coordinates \eqref{EF}, and also compare
it with the analogous Ludwig-Edgar solution. They found, among other
things, that both metrics are similar in a sense that both are Ricci-simple
but dissimilar with respect to belonging to pseudosymmetric or
semisymmetric manifolds. Note also that examination of the structure of the energy-momentum tensor for background radiation in 
\cite{Shaikh2019} is in concordance with our approach (cf. (\ref{enden}) as an application of the Einstein tensor).}.

\section{Outlook on the problem of variable mass in general relativity}%
\label{sec:outlook_on_the_problem_of_variable_masses_in_general_relativity}

The Vaidya problem describes the behavior of radial emission
background radiated from/onto the central non-rotating mass. For this reason, strictly
speaking, the right side of the Einstein equations must contain, apart from the
emission, a massive source, which is absent in explicit form in the Vaidya approach. The situation resembles the transfer from the Laplace equation to the Poisson equation with a point source in Newtonian gravity. In general relativistic case, this might lead to the following extension of the Vaidya problem:
\begin{equation}
	G_{\mu \nu} = \varkappa (\varepsilon k_\mu k_\nu + \rho u_\mu u_\nu),
\end{equation}
where $\rho(t,\vec{r}) = \delta(\vec{r}) M(t)$. 
 In this case the rate of loss of the
central mass $\dot{M}$ is determined by the radiation intensity, i.e.
\begin{equation}
	\Delta m(t) = \int_0^t  \int_V{\varepsilon(t',r)dV dt'}  = - \Delta M(t),
\end{equation}
where $m$ characterizes the background radiation. With such modification there is
the possibility from the start to adopt a definite law for the central mass time-dependence. Following
general Vaidya's algorithm, we may choose some static metric as a fiducial one. For it, we
believe that the most natural choice (dictated by the conformance to
observations) is the generalized Papapetrou-like exponential metric,
\begin{equation}
	d{s}^2 = e^{-2 \phi(t,R)} d{t}^2 - e^{2 \phi(t,R)} \left(
	d{R}^2 + R^2 d\Omega^2  \right) 
	\label{papa}
\end{equation}
with $ \phi(t,R)=M(t)/R$,
as opposed to 
Vaidya's preference of the Schwarzschild-like curvature coordinates
\eqref{curvcoord}. Naturally, using \eqref{papa} implies inclusion into consideration
of the scalar background, i.e. adding the scalar field energy-momentum tensor into the right side of the Einstein
equations:
\begin{equation}
	G_{\mu \nu} = \varkappa (\varepsilon k_\mu k_\nu + \rho u_\mu u_\nu -
	T_{\mu\nu}(\phi) ),
	\label{newV}
\end{equation}
as well as the related Klein-Gordon equation. Negative sign in (\ref{newV})
means that scalar field is taken in antiscalar regime \cite{Makukov2018}.

However, it is worth to note that the direct application of such algorithm may
encounter certain difficulties due to restrictions concerning the application of
distributions in the Einstein equations \cite{Geroch1987}. This problem might be
surmounted by applying a smoothing algorithm to the central mass
(i.e. to $\delta$-distribution) \cite{Makukov2016}. In this case, the singular Newtonian-type potential in the exponential metric \eqref{papa} may be transformed as
\begin{equation}
	\frac{M(t)}{R} \quad \Rightarrow \quad
	M(t)\frac{\text{erf}(\frac{R}{\sqrt{2\sigma}})}{R},
\end{equation}
where  $\sigma$ is the smoothing scale. In principle, such approach could allow to 
relate a definite mass-loss law to observed radial emission.

\section{Conclusion}
We have analyzed a peculiar feature of the Vaidya metric, which is that,
unlike in traditional methods, its derivation does not
require direct integration of the Einstein equations. The resulting algebraic
algorithm can also be successfully used, in particular, to extend the Vaidya approach to corpuscular emission
and, furthermore, to cosmological background taken into account.
 
The price for this unique possibility is that the solution of Einstein's
equations under consideration can {\em de facto} be obtained only up to an
unknown mass-function.
At the same time, both the mass-function $m(r, t)$ and the functional
(\ref{firstint}) must identically satisfy the non-trivial nonlinear partial
differential equation (\ref{u01}), or, on equal footing, its corollary
(\ref{A1}). As for the search for the explicit forms of $m (r, t)$, in any case
each concrete expression (if it exists) should be carefully tested with the full system of the Einstein equations.

In conclusion, the issue of the closed representation of variable masses in
arbitrary coordinates in general relativity, strictly speaking, remains open.
The Vaidya solution was the first significant step in this direction, but
implemented for a particular case of curvature coordinates. As an alternative, we propose an approach 
related to explicitly time-dependent mass as described in Outlook.

\begin{acknowledgements}
The work is partially supported within the grant No. AP08052312 of the Ministry of Education and Science of the Republic of Kazakhstan.
\end{acknowledgements}

\bibliographystyle{spphys}       
\bibliography{refs}   

\begin{thebibliography}{10}
\providecommand{\url}[1]{{#1}}
\providecommand{\urlprefix}{URL }
\expandafter\ifx\csname urlstyle\endcsname\relax
  \providecommand{\doi}[1]{DOI \discretionary{}{}{}#1}\else
  \providecommand{\doi}{DOI \discretionary{}{}{}\begingroup
  \urlstyle{rm}\Url}\fi

\bibitem{Padmanabhan2010}
T.~Padmanabhan, \emph{Gravitation: {Foundations} and {Frontiers}} (Cambridge
  University Press, 2010)

\bibitem{Vaidya1943}
P.~Vaidya, Current Science \textbf{12}, 183 (1943)

\bibitem{Vaidya1951}
P.C. Vaidya, Proceedings of the Indian Academy of Sciences - Section A
  \textbf{33}(5), 264 (1951).
\newblock \doi{10.1007@BF03173260}

\bibitem{Vaidya1953}
P.C. Vaidya, Nature \textbf{171}(4345), 260 (1953).
\newblock \doi{10.1038/171260a0}

\bibitem{Mychelkin1990a}
E.G. Mychelkin, A.A. Bekov.
\newblock The gravitational field of the variable mass on the cosmological de
  {Sitter} background (1990).
\newblock Preprint 90-05, Astrophysical Institute of the Academy of Sciences of
  the Kazakh SSR, Alma-Ata

\bibitem{Raychaudhuri1953}
A.K. Raychaudhuri, Zeitschrift f{\"u}r Physik \textbf{135}(2), 225 (1953).
\newblock \doi{10.1007/BF01333345}

\bibitem{Culetu2016}
H.~Culetu,   (2016).
\newblock \urlprefix\url{http://arxiv.org/abs/1612.06009}

\bibitem{Foster1972}
J.C. Foster, J.R. Ray, Journal of Mathematical Physics \textbf{13}(7), 979
  (1972).
\newblock \doi{10.1063/1.1666097}

\bibitem{Caban2006}
P.~Caban, J.~Rembieli{\'n}ski, K.A. Smoli{\'n}ski, Z.~Walczak, Foundations of
  Physics Letters \textbf{19}(6), 619 (2006).
\newblock \doi{10.1007/s10702-006-1015-4}

\bibitem{Ehrlich2015}
R.~Ehrlich, Astroparticle Physics \textbf{66}, 11 (2015).
\newblock \doi{10.1016/j.astropartphys.2014.12.011}

\bibitem{Mychelkin2015}
E.G. Mychelkin, M.A. Makukov, International Journal of Modern Physics D
  \textbf{24}, 1544025 (2015).
\newblock \doi{10.1142/S0218271815440253}

\bibitem{Makukov2016}
M.A. Makukov, E.G. Mychelkin, V.L. Saveliev, International Journal of Modern
  Physics: Conference Series \textbf{41}, 1660133 (2016).
\newblock \doi{10.1142/S2010194516601332}

\bibitem{Mallett1985}
R.L. Mallett, Physical Review D \textbf{31}(2), 416 (1985).
\newblock \doi{10.1103/PhysRevD.31.416}

\bibitem{Kozhanov1986b}
T.S. Kozhanov, E.G. Mychelkin, Proceedings of the Astrophysical Institute,
  Alma-Ata \textbf{45}, 85 (1986)

\bibitem{Tolman1949}
R.C. Tolman, \emph{Relativity, {Thermodynamics} and {Cosmology}} (Oxford at the
  Clarendon Press, London, 1949)

\bibitem{Zhuravsky1941a}
A.~Zhuravsky, \emph{Handbook of {Elliptical} {Functions}} (Academy of Science
  Press, Moscow, 1941)

\bibitem{Gupta1988}
Y.K. Gupta, S.~Gupta, General Relativity and Gravitation \textbf{20}(12), 1293
  (1988).
\newblock \doi{10.1007/BF00756054}

\bibitem{Fisher1948}
I.Z. Fisher, Zhurnal Experimental'noj i Teoreticheskoj Fiziki \textbf{18}, 636
  (1948)

\bibitem{Janis1968}
A.I. Janis, E.T. Newman, J.~Winicour, Physical Review Letters \textbf{20}(16),
  878 (1968).
\newblock \doi{10.1103/PhysRevLett.20.878}

\bibitem{Xanthopoulos1989}
B.C. Xanthopoulos, T.~Zannias, Physical Review D \textbf{40}(8), 2564 (1989).
\newblock \doi{10.1103/PhysRevD.40.2564}.
\newblock \urlprefix\url{https://link.aps.org/doi/10.1103/PhysRevD.40.2564}

\bibitem{Makukov2020}
M.~Makukov, E.~Mychelkin, Foundations of Physics  (2020).
\newblock \doi{10.1007/s10701-020-00384-y}

\bibitem{Dymnikova}
I.~Dymnikova, arXiv:gr-qc/0010016  (2000).
\newblock \urlprefix\url{http://arxiv.org/abs/gr-qc/0010016}

\bibitem{Kerr1963}
R.P. Kerr, Physical Review Letters \textbf{11}(5), 237 (1963).
\newblock \doi{10.1103/PhysRevLett.11.237}.
\newblock \urlprefix\url{https://link.aps.org/doi/10.1103/PhysRevLett.11.237}

\bibitem{Newman1965}
E.T. Newman, A.I. Janis, Journal of Mathematical Physics \textbf{6}(6), 915
  (1965).
\newblock \doi{10.1063/1.1704350}

\bibitem{Choptuik1993}
M.W. Choptuik, Physical Review Letters \textbf{70}(1), 9 (1993).
\newblock \doi{10.1103/PhysRevLett.70.9}

\bibitem{Shaikh2019}
A.A. Shaikh, H.~Kundu, J.~Sen, Indian Journal of Mathematics \textbf{61}, 41
  (2019)

\bibitem{Makukov2018}
M.A. Makukov, E.G. Mychelkin, Physical Review D \textbf{98}(6), 064050 (2018).
\newblock \doi{10.1103/PhysRevD.98.064050}

\bibitem{Geroch1987}
R.~Geroch, J.~Traschen, Physical Review D \textbf{36}(4), 1017 (1987).
\newblock \doi{10.1103/PhysRevD.36.1017}

\end{thebibliography}

\end{document}